\documentclass[twocolumn,aps,showpacs,pra,10pt]{revtex4-1}
\usepackage{amsmath}
\usepackage{amssymb}
\usepackage{graphicx}
\usepackage{hyperref}
\usepackage{color}

\begin{document}
\title{Spatial distribution of Spontaneous Parametric Down-Converted Photons for higher order Optical Vortices}
\author{Shashi Prabhakar,$^1$ Salla Gangi Reddy,$^1$ A Aadhi,$^1$ Ashok Kumar,$^2$ Chithrabhanu P,$^1$ G. K. Samanta$^1$ and R. P. Singh$^1$} 

\affiliation{$^1$Physical Research Laboratory, Navrangpura, Ahmedabad. 380009, India.}

\affiliation{$^2$Instituto de Fisica, Universidade de Sao Paulo, Sao Paulo, 66318, Brazil.}

\date{\today}

%%%%%%%%%%%%%%%%%%%%%%%%%%%%%%%%%%%%%%%%%%%%%%%%%%%%%%%%%%%%%%%%%%%%%%%%%%%%%%%%
%%%%%                          Abstract                                %%%%%%%%%
%%%%%%%%%%%%%%%%%%%%%%%%%%%%%%%%%%%%%%%%%%%%%%%%%%%%%%%%%%%%%%%%%%%%%%%%%%%%%%%%
\begin{abstract}
We make a source of entangled photons (SEP) using spontaneous parametric down-conversion (SPDC) in a non-linear crystal and study the spatial distribution of photon pairs obtained through the down-conversion of different modes of light including higher order vortices. We observe that for the Gaussian pump, the thickness of the SPDC ring varies linearly with the radius of pump beam. However, in case of vortex carrying beams, two concentric SPDC rings are formed for beams above a critical radius. The full width at half maximum (FWHM) of SPDC rings increase with increase in the order of optical vortex beams. The presence of a critical beam width for the vortices as well as the observed FWHM of the SPDC rings are supported with our numerical results.
\end{abstract}

%%%%%%%%%%%%%%%%%%%%%%%%%%%%%%%%%%%%%%%%%%%%%%%%%%%%%%%%%%%%%%%%%%%%%%%%%%%%%%%%
%%%%%                   PACS Codes                                      %%%%%%%%
%%%%%%%%%%%%%%%%%%%%%%%%%%%%%%%%%%%%%%%%%%%%%%%%%%%%%%%%%%%%%%%%%%%%%%%%%%%%%%%%
% 050.4865: Diffraction and gratings --> Optical vortices
% 190.4975: Nonlinear optics --> Parametric processes
% 42.65.-k: Nonlinear optics
% 42.65.Yj: Parametric oscillators and amplifiers, optical
\maketitle

%%%%%%%%%%%%%%%%%%%%%%%%%%%%%%%%%%%%%%%%%%%%%%%%%%%%%%%%%%%%%%%%%%%%%%%%%%%%%%%%
%%%%%                   Introduction                                    %%%%%%%%
%%%%%%%%%%%%%%%%%%%%%%%%%%%%%%%%%%%%%%%%%%%%%%%%%%%%%%%%%%%%%%%%%%%%%%%%%%%%%%%%
%%%%%%%%%%%%%%%%%%%%%%%%%%%%%%%%%%%%%%%%%%%%%%%%%%%%%%%%%%%%%%%%%%%%%%%%%%%%%%%%
%%%%%                   Introduction                                    %%%%%%%%
%%%%%%%%%%%%%%%%%%%%%%%%%%%%%%%%%%%%%%%%%%%%%%%%%%%%%%%%%%%%%%%%%%%%%%%%%%%%%%%%
\section{\label{sec:intro}Introduction}
The process of spontaneous parametric down-conversion (SPDC) has been used extensively for the generation of entangled photon pairs in many recent experiments. The purpose of these experiments range from Bell's inequality violation \cite{zhang_generation_2007} to the implementation of quantum information protocols \cite{kok_linear_2007}. In the process of SPDC, a laser pump beam photon interacts with second-order nonlinear $\chi^{(2)}$ crystal, gets annihilated and gives rise to the emission of two photons. These two photons are generated simultaneously and follow the laws of energy and momentum conservation. The phenomena of SPDC was first observed by Burnham and Weinberg \cite{burnham_observation_1970} and theoretically studied by Hong Ou and Mandel \cite{hong_theory_1985}.

The photon pairs generated through SPDC are entangled in the spatial degrees of freedom i.e. position-momentum entanglement \cite{edgar_imaging_2012} as well as entanglement in orbital angular momentum (OAM) \cite{molina-terriza_twisted_2007}. This OAM entanglement can be described by a multi-dimensional Hilbert space \cite{law_analysis_2004, mair_entanglement_2001, howell_realization_2004}, compared to the case of polarization entanglement which is limited to two dimensions only \cite{kwiat_new_1995}. These photon pairs have been found to be entangled in time-bin also \cite{simon_creating_2005}.

Optical vortices (OV) carry a dark core in a bright background \cite{allen_optical_2003}. If there is a phase change of $2\pi l$ around the point of darkness, it is called a vortex of topological charge $l$, where $l$ is an integer. The sense of rotation determines the sign of topological charge of the vortex. A beam with such a phase structure has a helical wavefront and, therefore, carries an OAM of $l\hslash$ per photon \cite{soskin_topological_1997} for a vortex of topological charge $l$. These beams have found a variety of applications, such as optical trapping of atoms \cite{kuga_novel_1997}, optical tweezing and spanning \cite{grier_revolution_2003}, optical communication \cite{bozinovic_terabit-scale_2013}, imaging \cite{brunet_transverse_2010}, and quantum information and computation \cite{mair_entanglement_2001}.

For any application of entangled photons generated through the SPDC, it is important to know the spatial distribution of photons arising from the SPDC process. For the Gaussian pump beam, the spatial distribution of SPDC photons has already been reported \cite{di_lorenzo_pires_type-i_2011, ramirez-alarcon_effects_2013, grayson_spatial_1994, cruz-ramirez_observation_2012}. However, for photons generated by pumping with higher order vortices, it has not been reported so far. Although, the phase-matching by optical vortex pump beam h as been studied theoretically by Pittman et.al. \cite{pittman_two-photon_1996}.

With the availability of low noise and high quantum-efficiency electron-multiplying CCDs (EMCCD), the experiments with low photon level imaging has become possible \cite{fickler_real-time_2013}. To observe the shape of the SPDC ring formed by the Gaussian as well as optical vortex beams, we have carried out experimental studies using EMCCD. The observed experimental results are supported with our numerical results. The theory regarding the SPDC has been discussed in section \ref{sec:theory}, experiments performed in section \ref{sec:expsetup} and results in section \ref{sec:result}. Finally we conclude in section \ref{sec:conclusion}.

%%%%%%%%%%%%%%%%%%%%%%%%%%%%%%%%%%%%%%%%%%%%%%%%%%%%%%%%%%%%%%%%%%%%%%%%%%%%%%%%
%%%%%                   THEORY                                          %%%%%%%%
%%%%%%%%%%%%%%%%%%%%%%%%%%%%%%%%%%%%%%%%%%%%%%%%%%%%%%%%%%%%%%%%%%%%%%%%%%%%%%%%
\section{\label{sec:theory}Theory}
The intensity distribution of an optical vortex of order $l$ can be written as
\begin{equation}\label{eq:gaussian}
I_l(x,y)=I_0(x^2+y^2)^{|l|}\exp\left(-\frac{x^2+y^2}{\sigma^2}\right),
\end{equation}
where $\sigma$ is the beam radius of host beam, $I_0$ is the maximum intensity in the bright ring. Clearly, Eq. \ref{eq:gaussian} shows that the Gaussian beam is a special case of optical vortex with $l=0$.

The nonlinear effects in crystals have been exploited in a number of applications such as frequency doubling, optical parametric oscillation and the SPDC \cite{boyd_nonlinear_2003}. When a nonlinear crystal, for example Beta-Barium Borate (BBO), with non-zero second order electric susceptibility ($\chi^{(2)}$) is pumped by an intense laser, a \textit{pump} photon (frequency $\omega_p$ and wave-vector $\mathbf{K}_p$) splits into a photon pair called \textit{signal} and \textit{idler}. The energy and momentum conservation provide us with
\begin{eqnarray}
  \hbar\omega_p &=&\hbar\omega_s+\hbar\omega_i, \label{eq:energy_conservation}\\
  \mathbf{K_p} &=&\mathbf{K_s} +\mathbf{K_i}, \label{eq:phase_matching}
\end{eqnarray}
where suffices $s$ and $i$ denote signal and idler photons respectively. The phase matching is determined by the frequency of pump laser beam, and the orientation of crystal optic axis with respect to the pump. Equation \ref{eq:energy_conservation} can be simplified as
\begin{equation}
\frac{1}{\lambda_p} = \frac{1}{\lambda_s} + \frac{1}{\lambda_i}, \label{eq:energy}
\end{equation}
where $\lambda_p$, $\lambda_s$ and $\lambda_i$ denote wavelengths of pump, signal and idler photons respectively. We have considered e $\rightarrow$ o + o type (e: extraordinary, o: ordinary) interaction. Hence, Eq. \ref{eq:phase_matching} can be written as
\begin{eqnarray}
\frac{2\pi n_{e}(\lambda_p,\Theta)}{\lambda_p}&=&\frac{2\pi n_o(\lambda_s)}{\lambda_s}\cos(\phi_s)+\frac{2\pi n_o(\lambda_i)}{\lambda_i}\cos(\phi_i) \label{eq:momentum1} \\
\frac{2\pi n_o(\lambda_s)}{\lambda_s}\sin(\phi_s)&=&\frac{2\pi n_o(\lambda_i)}{\lambda_i}\sin(\phi_i)
\label{eq:momentum2}
\end{eqnarray}
where $\phi_s$ is the angle between $\mathbf{K_p}$ and $\mathbf{K_s}$, $\phi_i$ is the angle between $\mathbf{K_p}$ and $\mathbf{K_i}$ and $\Theta$ is the direction of optic axis with respect to $\mathbf{K_p}$. $n_e(\lambda_p,\Theta)$ and $n_o(\lambda_{s,i})$ are the extraordinary and ordinary refractive indices for respective wavelengths. They are obtained from the Sellmeier equations \cite{boyd_nonlinear_2003} and for the BBO crystal used in the experiment can be written as
\begin{eqnarray}
n_o(\lambda) &=& \sqrt{2.7359 + \frac{0.01878}{\lambda^2 - 0.01822} - 0.01354\lambda^2 } \\
n_e(\lambda) &=& \sqrt{2.3753 + \frac{0.01224}{\lambda^2 - 0.01667} - 0.01516\lambda^2 } \\
n_e(\lambda,\Theta) &=& n_o(\lambda) \sqrt{\frac{1+\tan(\Theta)^2}{1+\left[\frac{n_o(\lambda)}{n_e(\lambda)}\times\tan(\Theta)\right]^2}} \label{fig:fromphasematching}
\end{eqnarray}
where $\lambda$ is in $\mu$m.

\begin{figure}[b]
  \begin{center}
    \includegraphics[width=7cm]{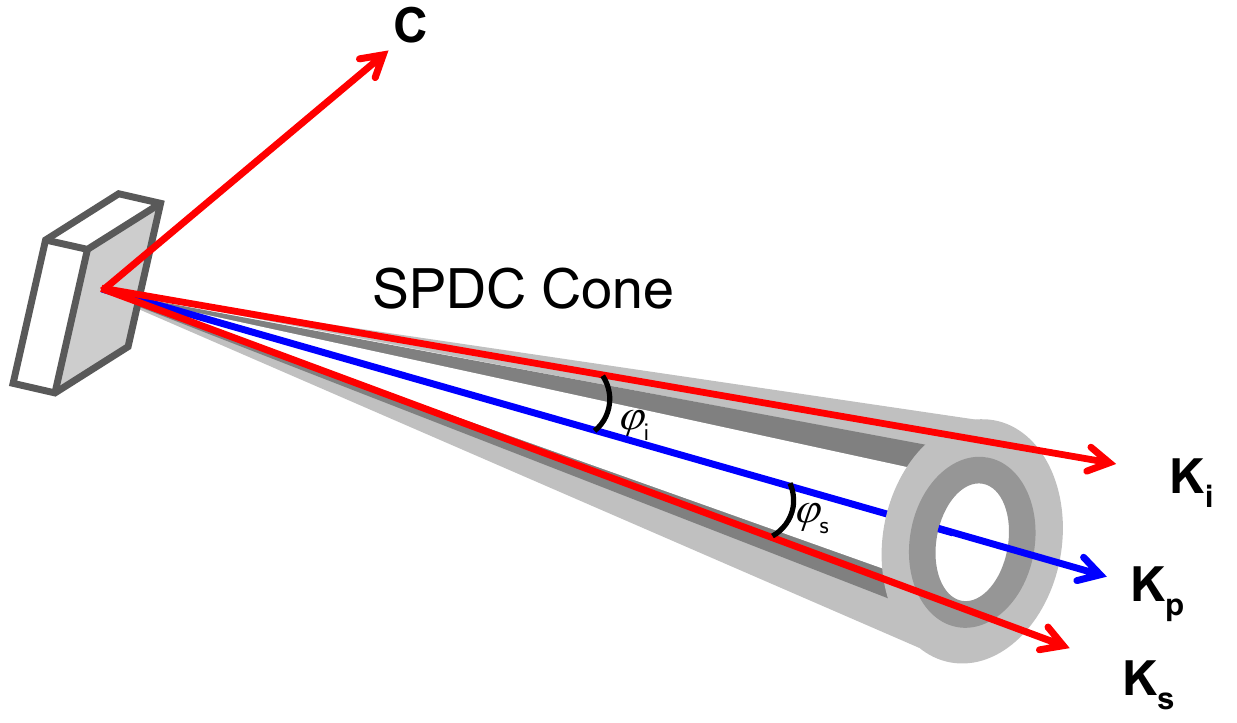}
    \caption{(Color online) Sketch diagram for the SPDC ring emission after passing the pump beam through the BBO crystal. Light and dark gray levels represent generation of idler and signal photon SPDC rings respectively.}
    \label{fig:spdc_process}
  \end{center}
\end{figure}

In Fig. \ref{fig:spdc_process}, we have given a sketch of the SPDC photon pair generation in non-collinear type-I SPDC process. $\mathbf{C}$ denotes the crystal optic axis. The angular separation between $\mathbf{K_p}$ and $\mathbf{K_s}$ is due to energy and phase-matching conditions (Eq. \ref{eq:momentum1} and \ref{eq:momentum2}) required for the SPDC process. We have also shown generation of a pair of signal and idler photons and formation of the ring centered around $\mathbf{K_p}$. In the present case, we have assumed that the pump beam has same horizontal and vertical widths.

We have used a negative-uniaxial BBO crystal with non-linear coefficient $d_{\rm eff}$ = 2.00 pm/V, thickness 5 mm and optic axis $\Theta$ = 29.7$^\circ$. The pump beam with wavelength $\lambda_p$ = 405 nm is incident normal to the crystal. We plan to study the degenerate or near-degenerate case in which the signal and idler photons have almost same wavelength $\lambda_{s,i}$ = 810$\pm$5 nm. The wavelength for down-converted photons are chosen from the interference filters (IF) used in the experiment. With these experimental parameters, Eqs. \ref{eq:momentum1} and \ref{eq:momentum2} have been solved to determine $\phi_s$ and $\phi_i$ by Runge-Kutta (RK) method for a particular value of $\lambda_s$ and $\lambda_i$ satisfied by Eq. \ref{eq:energy}.

Numerical simulations have been performed by first considering a particular value of $\lambda_s$ and $\lambda_i$. Angles $\phi_s$ and $\phi_i$ are evaluated using RK method for chosen $\lambda_s$, $\lambda_i$ and experimental parameters. The signal and idler photons are generated in cones having half-opening angle $\phi_s$ and $\phi_i$ as represented in Fig. \ref{fig:spdc_process} and appear as two rings on the detector plane. The center of these rings are concentric with the pump beam. Now, consider a single point on the intensity distribution of pump falling on the crystal. The stream of single photons passing through the chosen point generates SPDC rings whose radius depends on the distance between crystal and EMCCD. The intensity of the rings are proportional to the intensity at the selected point. The rings corresponding to signal and idler photons are then added to obtain the SPDC ring for the pump photons. In the similar way, rings for all other points of pump intensity distribution are obtained and added. The obtained spatial distribution will depend on the shape and size of the pump beam. This will provide the SPDC ring for $\lambda_s$ and $\lambda_i$. The contributions due to whole wave-length range (810$\pm$5 nm)allowed by the IF has been considered to obtain the resultant spatial distribution of SPDC ring.

%%%%%%%%%%%%%%%%%%%%%%%%%%%%%%%%%%%%%%%%%%%%%%%%%%%%%%%%%%%%%%%%%%%%%%%%%%%%%%%%
%%%%%                   Theory and Methods                              %%%%%%%%
%%%%%%%%%%%%%%%%%%%%%%%%%%%%%%%%%%%%%%%%%%%%%%%%%%%%%%%%%%%%%%%%%%%%%%%%%%%%%%%%

\section{\label{sec:expsetup}Experimental Setup}
\begin{figure}[h]
  \begin{center}
    \includegraphics[width=7cm]{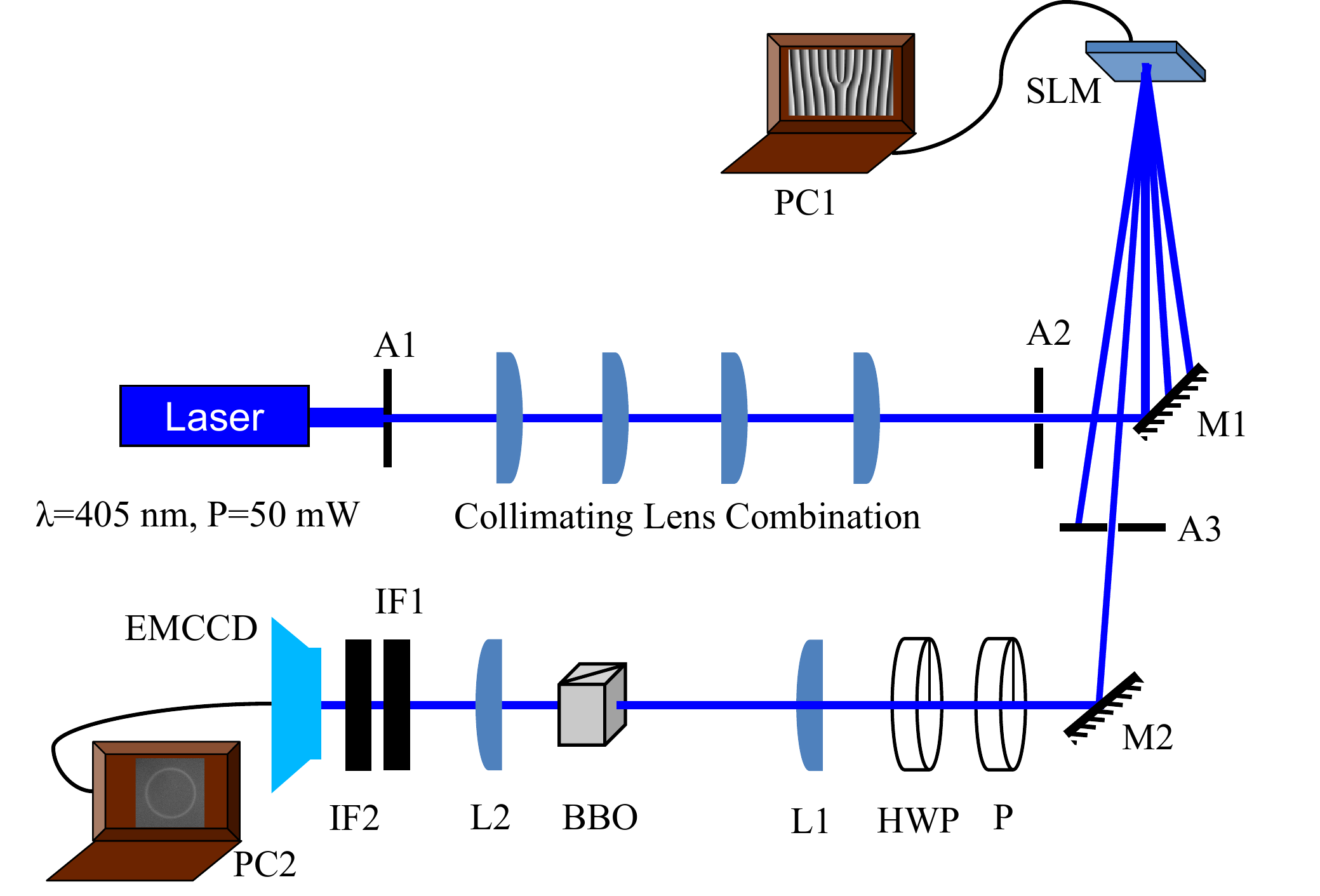}
    \caption{(Color online) Experimental setup for the study of SPDC photon pair distribution with an optical vortex as pump beam.}
    \label{fig:vort_expsetup}
  \end{center}
\end{figure}

The experimental set-up to study the SPDC photon distribution generated by the Gaussian as well as the optical vortex pump beam is shown in Fig. \ref{fig:vort_expsetup}. The astigmatism of the diode laser (RGBLase 405 nm, 50 mW) have been removed by using a combination of lenses. The collimated beam is then sent to a spatial light modulator (SLM) (Hamamatsu LCOS SLM X-10468-05), which is interfaced with computer (PC1).	 Blazed holograms have been used to generate OV with higher power in the first diffraction order \cite{bryngdahl_formation_1970}. The first diffracted order is selected with an aperture A3. Polarizer (P) and half-wave plate (HWP) are used to select and rotate the polarization of pump beam respectively. BBO crystal (6$\times$6$\times$5 mm$^3$) with optic axis at 29.7$^\circ$ is used for the parametric down-conversion. As the size of OV beams of higher order becomes bigger than the size of the crystal, we have used a lens L1 ($f$=15 cm) to loosely focus the vortex beam on the crystal. The BBO crystal is mounted on a rotation stage, so that phase-matching angle can be achieved by rotating the crystal along its optic axis. After achieving phase-matching, the crystal remains unaltered for all the observations.

When phase-matched, the output cone makes half angle of $\sim4^{\circ}$ with pump direction $\mathbf{K_p}$. The BBO crystal is kept in such a way that it can down-convert only vertically polarized light. Therefore, when angle of the HWP is 0$^{\circ}$ (45$^{\circ}$), then we will get (not get) down-converted photons. Image of the down-converted ring is recorded by Andor iXon$_3$ EMCCD camera using an imaging lens of focal length 5 cm. We have used the EMCCD in background correction mode. In this mode, background is obtained when $\lambda/2$ plate is at 45$^\circ$ and signal is obtained when $\lambda/2$ plate is at 0$^\circ$. The central bright spot in experimental observations show the unfiltered pump beam. This could not be subtracted while subtracting the background due to shift in its position during the rotation of HWP from 45$^\circ$ to 0$^\circ$. The interference filters IF1 and IF2 pass only the down-converted photons of wave-length 810$\pm$5 nm and block the pump photons of wave-length 405 nm. Two interference filters have been used to reduce the pump photons as much as possible.

The power of 405 nm laser falling on the BBO crystal was 2 mW. EMCCD was operated at -80$^\circ$C. Further, we have taken images by accumulating 50 frames exposure time of 1 s. We have used the complete 512$\times$512 pixels of the camera. The readout rate was set at 1 MHz 16-bit. Since the observed SPDC rings were sufficiently intense, we have not enabled the electron-multiplication gain.

The size of pump beam has been measured by imaging the beam at the position of crystal with Point-Grey (FL2-20S4C) CCD camera. The images obtained from the CCD camera are read in Matlab for further processing. The 2-D curve fitting is used to obtain the best-fit intensity distribution obtained by Eq. \ref{eq:gaussian} that provides us with beam-width of the pump ($\sigma_{\rm pump}$). For our numerical calculations, we have used the best-fit value of $\sigma_{\rm pump}$ obtained experimentally.

%%%%%%%%%%%%%%%%%%%%%%%%%%%%%%%%%%%%%%%%%%%%%%%%%%%%%%%%%%%%%%%%%%%%%%%%%%%%%%%%
%%%%%                   Result                                          %%%%%%%%
%%%%%%%%%%%%%%%%%%%%%%%%%%%%%%%%%%%%%%%%%%%%%%%%%%%%%%%%%%%%%%%%%%%%%%%%%%%%%%%%
\section{\label{sec:result}Result and Discussion}
The objective of the experimental work is to characterize the spatial distribution of degenerate SPDC photon pairs produced by higher order vortices and verify the results obtained with numerical calculations. Before pumping the nonlinear crystal with high order vortices, we study the distribution of SPDC photons generated by the Gaussian beam of different widths.

\begin{figure}[t]
  \begin{center}
    \includegraphics[width=7cm]{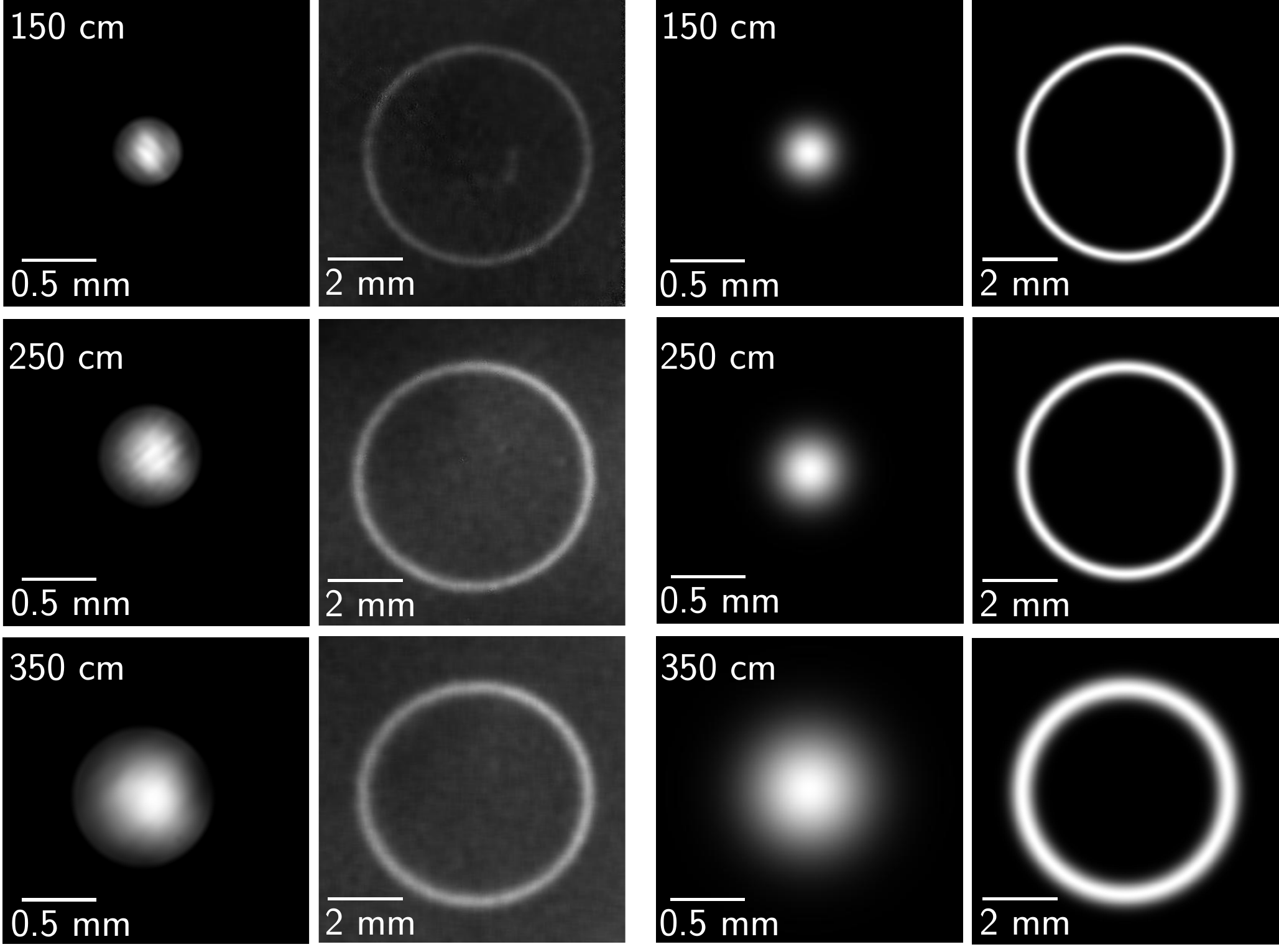}
    \caption{Experimental (left) and Numerical (right) are Gaussian pump and their corresponding SPDC rings recorded at distances 150 cm, 250 cm and 350 cm from the SLM.}
    \label{spdc_gauss}
  \end{center}
\end{figure}

To make a comparison of spatial distribution of down-converted photons due to the Gaussian and the vortex beams, the Gaussian beam is generated using the SLM by transferring the blazed grating hologram of topological charge 0 to the SLM. To vary the width of the Gaussian beam, we have used the beam at different propagation distances from the SLM (150 cm to 350 cm in the steps of 50 cms). As the size of beam was lower than the aperture of the crystal at 350 cm from the SLM, lens (L1) was not used. The experimentally and numerically obtained SPDC rings are shown in Fig. \ref{spdc_gauss}. We observe an increase in thickness of the SPDC ring as the pump beam size increases.

\begin{figure}[b]
  \begin{center}
    \includegraphics[width=7cm]{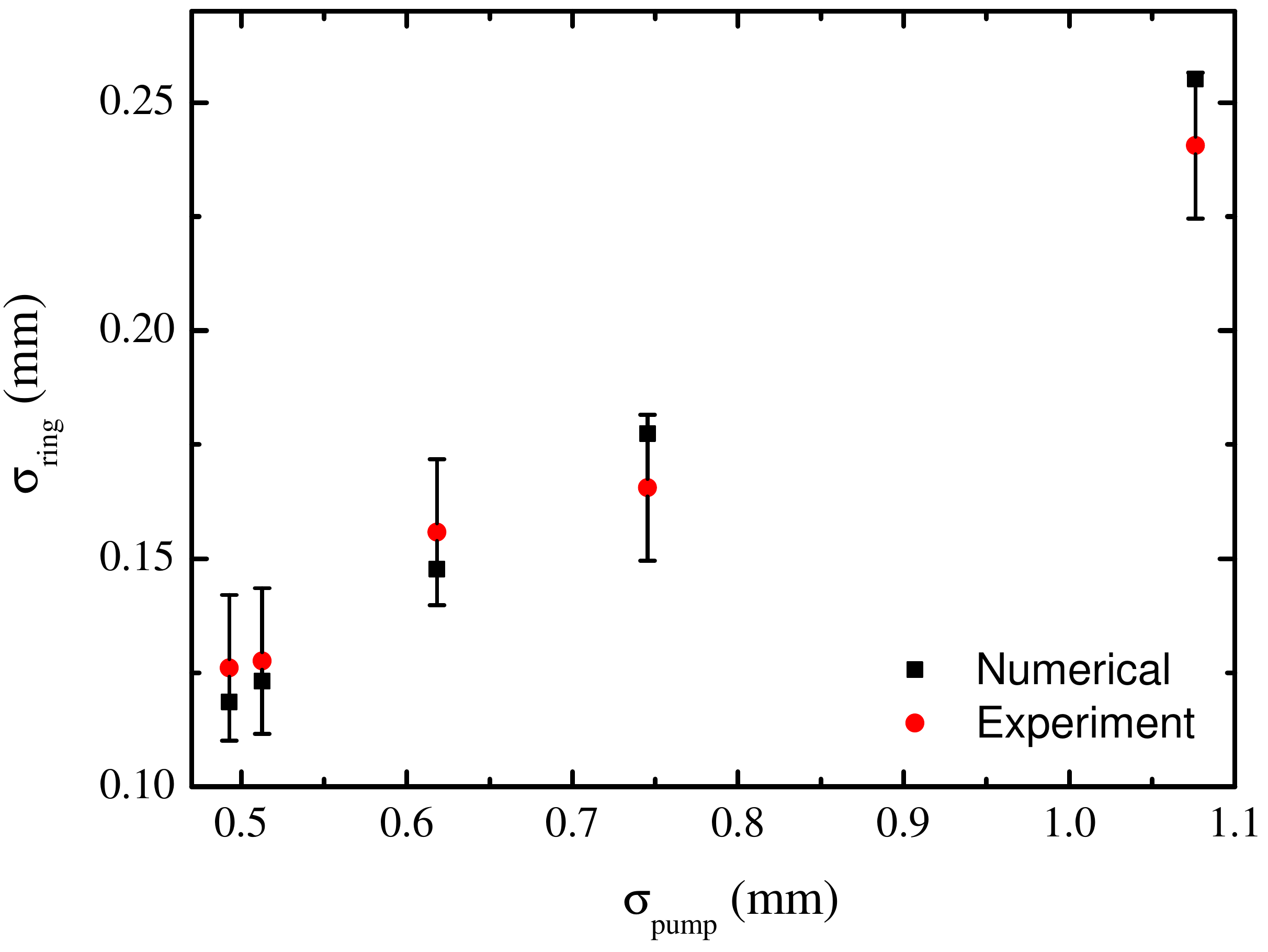}
    \caption{(Color online) Variation of $\sigma_{\rm ring}$ with $\sigma_{\rm pump}$. The curve shows the linear variation in thickness of the SPDC ring with the beam-width of Gaussian pump beam.}
    \label{gauss_graph}
  \end{center}
\end{figure}

To obtain a quantitative variation of SPDC ring, we use the line profile through the SPDC rings along their center. Numerically, we have observed that the SPDC ring fits with a Gaussian function. To calculate the width of SPDC ring ($\sigma_{\rm ring}$), the profiles obtained are fitted with a Gaussian function as in Eq. \ref{eq:gaussian} for $l$=0. The variation of thickness of the SPDC rings with the size of the pump beam is shown in Fig. \ref{gauss_graph}. Numerical and experimental results are found to be in good agreement with each other. We find our results are similar to the one obtained earlier \cite{ramirez-alarcon_effects_2013}.

\begin{figure}[h]
  \begin{center}
    \includegraphics[width=8cm]{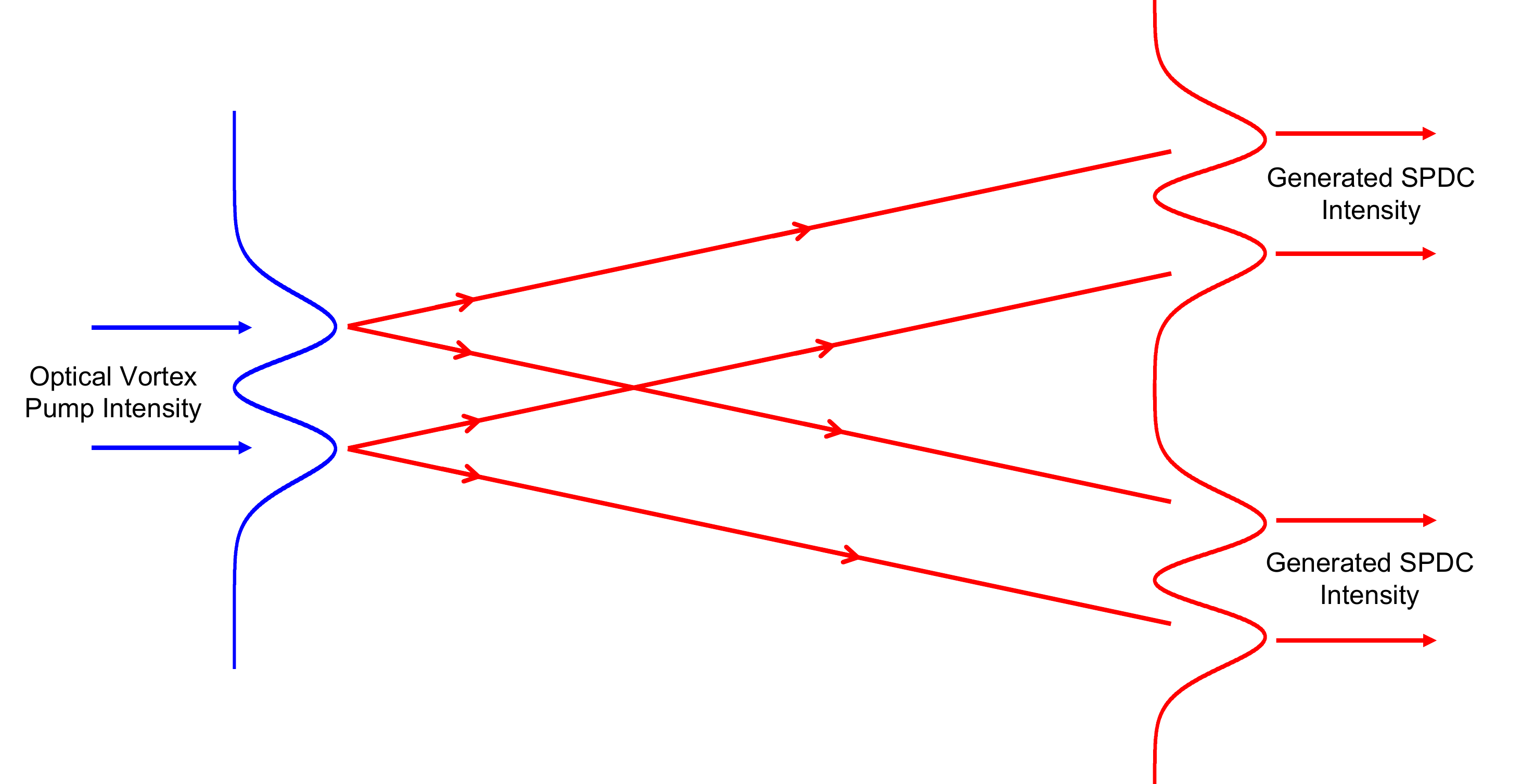}
    \caption{Schematic diagram for the generation of two rings when pumped with optical votex beams.}
    \label{fig:schematic_dual}
  \end{center}
\end{figure}

\begin{figure}[h]
  \begin{center}
    \includegraphics[width=12cm, angle=-90]{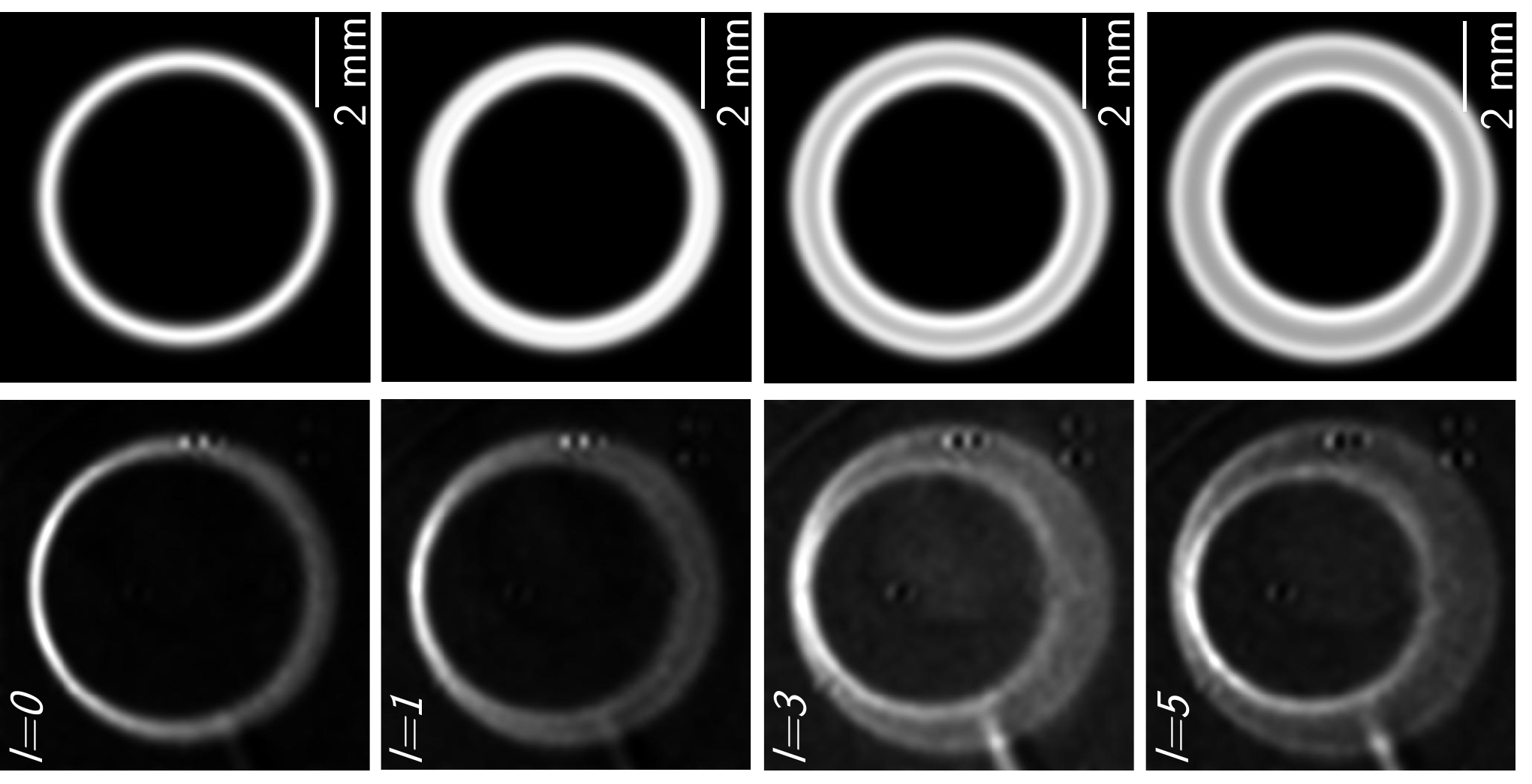}
    \caption{Experimental (left) and Numerical (right) SPDC rings due to an optical vortex pump beam for order 0, 1, 3 and 5. Spot in the center of experimental images correspond to the unfiltered pump beam.}
    \label{spdc_vort}
  \end{center}
\end{figure}

Figure \ref{fig:schematic_dual} shows the generation of two rings when the BBO crystal is pumped with OV. The blue and red lines show the intensity distribution of the pump and the SPDC photons. As the size of optical vortex goes beyond the aperture of BBO crystal, we have used lens (L1) to loosely focus it. It has been observed that the SPDC ring due to optical vortex forms two concentric bright rings with non-zero intensity in middle. The SPDC rings due to optical vortices are shown in Fig. \ref{spdc_vort}. From these images, we can observe the increase in thickness of the SPDC ring.

\begin{figure}[h]
  \begin{center}
    \includegraphics[width=7cm]{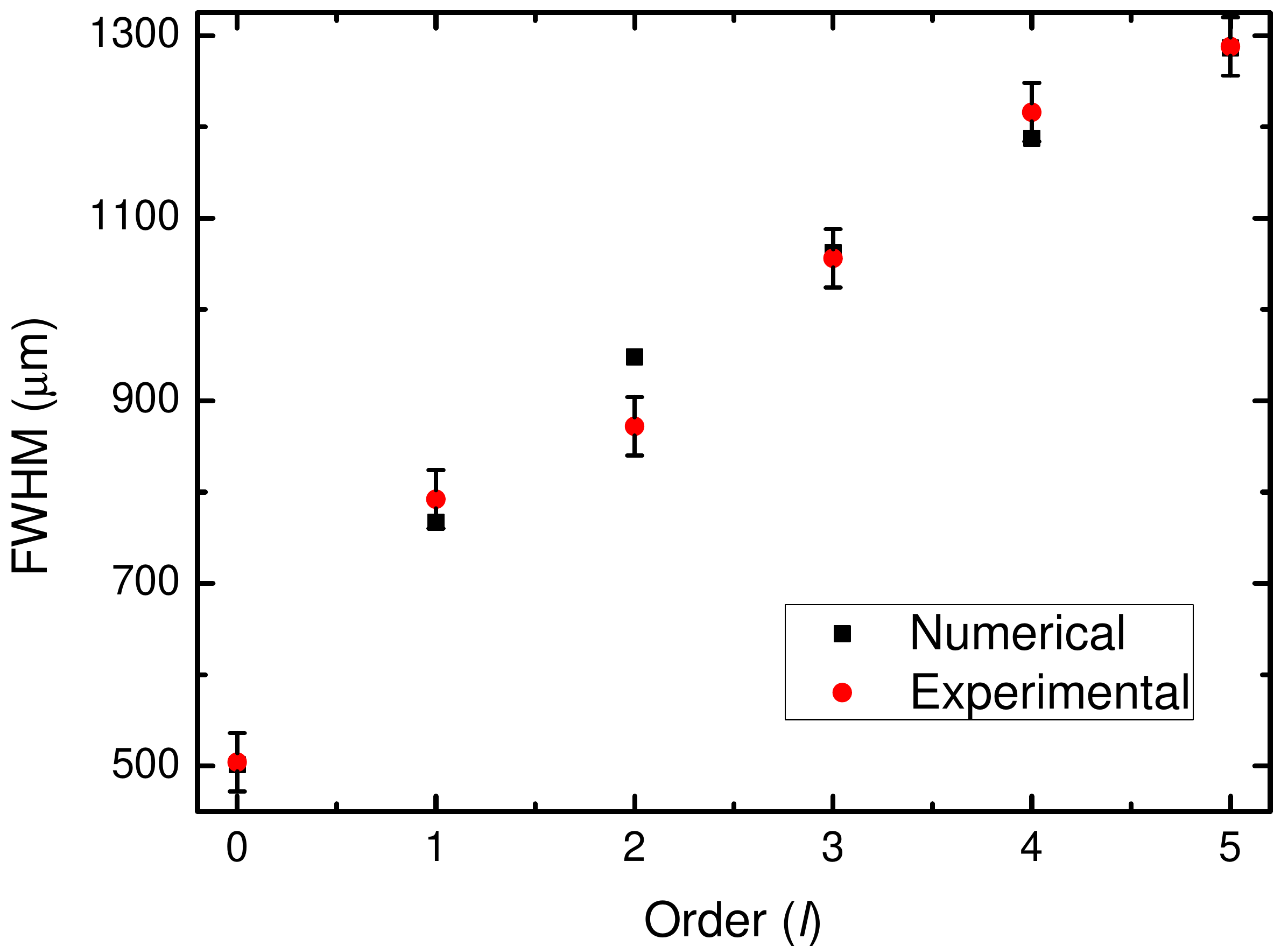}
    \caption{(Color online) Variation of FWHM of SPDC ring for optical vortex pump beams with the order of optical vortices.}
    \label{vortex_graph}
  \end{center}
\end{figure}

With the increase in topological charge of vortices, the full width at half maximum (FWHM) of the ring increases. The separation between the inner and the outer ring also  increases with the increase in order as shown in Fig. \ref{vortex_graph}. However, we can observe the asymmetry caused due to the crystal length. This asymmetry  arises due to the longitudinal phase matching and depends on nonlinear crystal properties of the crystal, including crystal length \cite{ramirez-alarcon_effects_2013}. This is one of the factors which affects the selection of entangled photons and consequently the total coincidence counts.

\begin{figure}[h]
  \begin{center}
    \includegraphics[width=7cm]{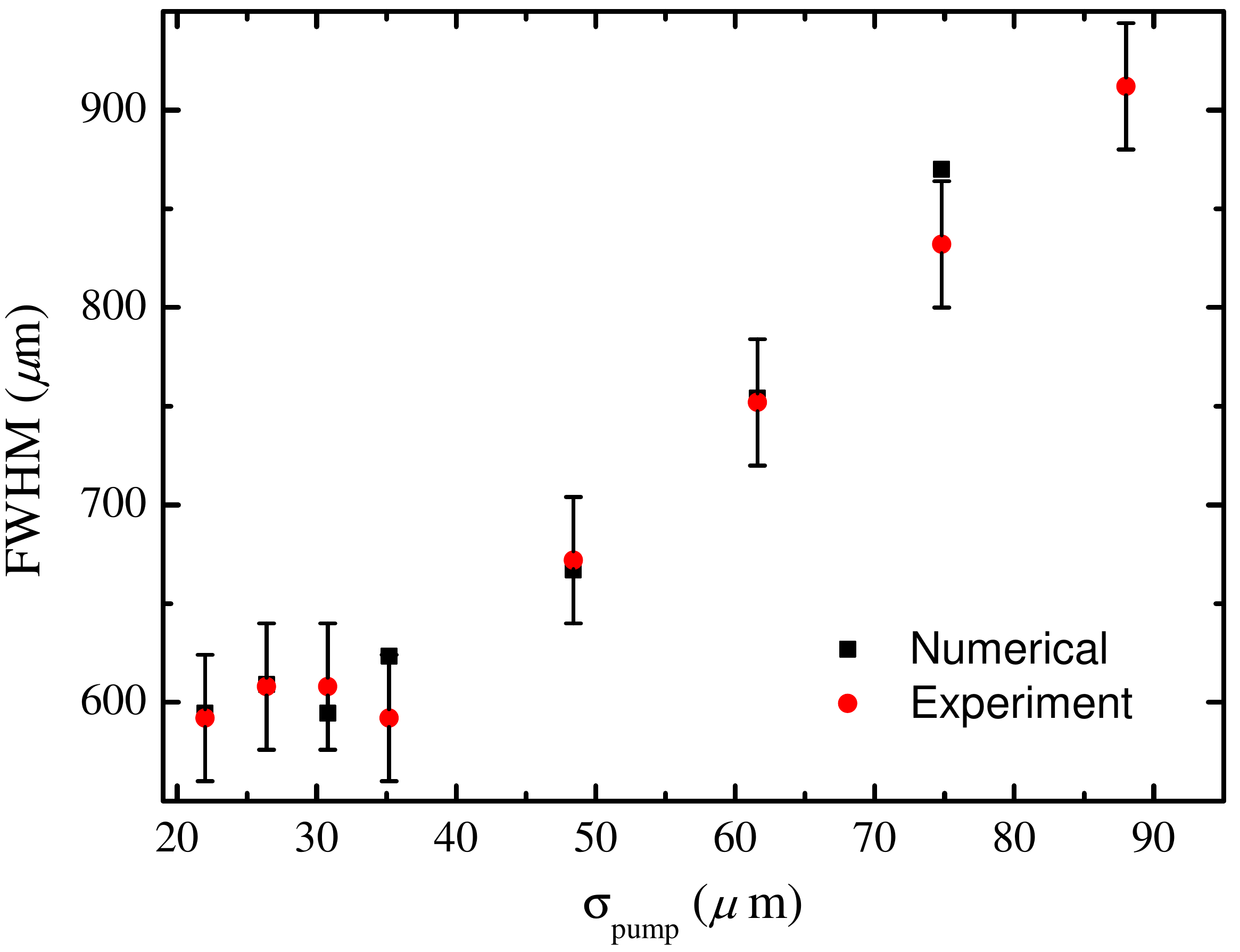}
    \caption{(Color online) Variation of FWHM of SPDC ring for optical vortex pump beam of order $l$=2 with beam width of host beam.}
    \label{vortex_lowerbound}
  \end{center}
\end{figure}

We have also observed that if $\sigma_{\rm pump}$ is lower than a particular value for the OV of topological charge $l$, then there will not be any change in FWHM. This variation has been studied by varying $\sigma_{\rm pump}$ and keeping the order $l$ fixed. We have observed that the FWHM of the ring starts increasing only when $\sigma_{\rm pump}$ is more than a particular beam size, called critical beam size. The variation of FWHM of SPDC ring for order $l$=2 with $\sigma_{\rm pump}$ is shown in Fig. \ref{vortex_lowerbound}. In case of OV, the numerical and experimental results are in good agreement with each other.

%%%%%%%%%%%%%%%%%%%%%%%%%%%%%%%%%%%%%%%%%%%%%%%%%%%%%%%%%%%%%%%%%%%%%%%%%%%%%%%%
%%%%%                   Conclusions and Discussion                      %%%%%%%%
%%%%%%%%%%%%%%%%%%%%%%%%%%%%%%%%%%%%%%%%%%%%%%%%%%%%%%%%%%%%%%%%%%%%%%%%%%%%%%%%
\section{\label{sec:conclusion}Conclusion}
The spatial distribution of entangled photons generated by non-linear crystal is of importance in the field of quantum information and quantum computation. We have observed a linear increase in thickness of the SPDC ring with beam radius of the pump.

We have also observed the formation of two concentric SPDC rings if the crystal is pumped with optical vortex beams. One of the reasons for generation of two rings is the dark core of optical vortex i.e. specific intensity distribution of the vortex. The numerical and experimental widths of the SPDC ring are in good agreement with each other. The formation of two rings takes place when the pump beam size is more than the critical beam size. These observations would be useful in the experiments to maximize the coincidence counts. Physically, the broadened SPDC is a consequence of the greater spread of pump transverse wave-vectors, resulting in phase matching for a greater spread of signal and idler transverse wave-vectors.

%%%%%%%%%%%%%%%%%%%%%%%%%%%%%%%%%%%%%%%%%%%%%%%%%%%%%%%%%%%%%%%%%%%%%%%%%%%%%%%%
%%%%%                   Acknowdedgements                                %%%%%%%%
%%%%%%%%%%%%%%%%%%%%%%%%%%%%%%%%%%%%%%%%%%%%%%%%%%%%%%%%%%%%%%%%%%%%%%%%%%%%%%%%
\section*{Acknowledgments}
Authors wish to acknowledge Dr. A. K. Jha for fruitful discussion and experimental techniques. The numerical calculations reported in this article have been performed on a 3 TeraFlop high-performance cluster at Physical Research Laboratory (PRL), Ahmedabad, India.

\end{document}